\documentstyle[emulateapj,times]{article}

\begin{document}

\def\insertplot#1#2#3#4#5#6#7{
\vskip 10pt\nobreak\hbox to \hsize{\hss\dimen0=#3in\hbox to #6\dimen0{%
\dimen0=#2in\vbox to #6\dimen0{\vss
\special{ps: plotfile #1}
\special{ps::[end]
  PGPLOT restore
}
}\hss}\hss}\vskip 10pt}

\slugcomment{\centerline{\bf \hfil accepted by ApJL; 10/24/00}}

\title{Periodic Photometric Variability in the Becklin-Neugebauer Object}

\author{Lynne A. Hillenbrand and John M. Carpenter}
\affil{California Institute of Technology,
Department of Astronomy, MS 105-24, \\ Pasadena, CA 91125;
email: lah@astro.caltech.edu, jmc@astro.caltech.edu}

\author{M.F. Skrutskie}
\affil{University of Massachusetts, Department of Astronomy, \\
Amherst, MA 01003; email: skrutski@north.astro.umass.edu}

\begin{abstract}

The Becklin-Neugebauer (BN) object in the Orion Nebula Cluster (ONC)
is a well-studied optically invisible, infrared-bright young stellar 
object, thought to be an intermediate-mass protostar.  
We report here that BN exhibited nearly-sinusoidal periodic variability 
at the near-infrared H- and K$_s$-bands during a one month observing campaign
in 2000 March/April. The period was 8.28 days and the peak-to-peak amplitude 
$\sim$0.2 mag.  Plausible mechanisms for producing the observed variability 
characteristics are explored.

\end{abstract}

\keywords{infrared: stars --- stars: individual (BN) --- stars: pre-main sequence --- stars: variables -\clearpage}

\section{Introduction}

The BN object was revealed during early raster scanning of the Orion Nebula
region at 2$\micron$, and immediately recognized as 
a candidate protostellar object (Becklin \& Neugebauer 1967).  
BN is the brightest at near-infrared wavelengths of a group 
of $\sim$20 intermediate- and high-mass young stars and protostars in OMC-1.  
These objects were discovered at mid-infrared (Rieke, Low, \& Kleinman 1973;
Downes et al. 1981; Lonsdale et al. 1982; Dougados et al. 1993; 
Gezari, Backman, \& Werner 1998), radio (Churchwell et al. 1987; Felli et al. 1993; 
Menten \& Reid 1995), and x-ray (Garmire et al. 2000) wavelengths.  
The total luminosity emanating from the embedded cluster 
is $\sim$10$^5 L_{\odot}$.  In addition to its luminous 
point sources, the BN region is also notable as the source of the spectacular
H$_2$ ``fingers'' or ``bullets'' (Allen \& Burton 1993) which extend 
several arcminutes to the northwest and southeast.  

BN itself is extinguished by A$_V$ = 17 mag and has luminosity 
2500 L$_{\odot}$ (Gezari et al.) corresponding to
a main sequence B3-B4 star.  It was the first of a still small class
of young, mostly luminous, stars with the 2$\micron$ CO bandheads in emission 
(Scoville et al. 1979, 1983).  These $\Delta\nu$=2 transitions are thought to 
arise from collisional or shock excitation in a hot, dense region, 
perhaps the inner part of a circumstellar disk or wind.
BN also has relatively strong H$_2$ (Scoville et al. 1983) and weak 
Hu, Pf, and Br hydrogen recombination lines (e.g. Bunn, Hoare, \& Drew 1995). 
Broad absorptions at 
3.3$\micron$ and 10$\micron$ are due to circumstellar ice and dust 
(Gillett \& Forrest 1973).

\section{Photometric Monitoring Observations}

Time series photometry at J, H, and K$_s$ 
was acquired using the 2MASS southern telescope at Cerro Tololo
during gaps in right ascension not otherwise utilized near 
completion of the 2MASS survey. 
As described in Carpenter, Hillenbrand \& Skrutskie (2001), 29 sets of 
photometry were obtained over an area 
$<\sim0.84^\circ\times6.0^\circ$ 
centered on the Trapezium region of the ONC. 
Observations were conducted on nearly
a nightly basis between 2000, March 4 and April 8 with BN 
observed on 28 of these nights. 
In addition to these specially scheduled observations, 
BN was observed twice during normal 
2MASS survey operations on 1998, March 19 and 2000, February 6.

As detailed in the 2MASS Explanatory Supplement (Cutri et al. 2000),
the image data consist of doubly-correlated
differences of two NICMOS readouts separated by the 1.3s
frame integration time.  The first readout occurs
51ms after reset and independently provides a short
integration to recover unsaturated images of bright
(5-9 mag) stars.  Each position on the sky is observed 6 times 
in this manner as the telescope scans in declination.
For BN, the photometric measurements at K$_s$-band are derived from the 51ms 
integrations since the source saturates in the 1.3s images. 
Magnitudes are obtained 
using aperture photometry with an aperture radius of 4$''$ and a sky annulus 
extending radially from 24$''$-30$''$. The final magnitude is the mean of the 
six aperture magnitudes, and the photometric uncertainty is the standard 
deviation of the mean of the six measurements. At H-band, BN is faint enough 
that the magnitudes normally would be estimated with Point 
Spread Function (PSF) fitting on the 1.3s images. However, 
since BN is located on an extended plateau of bright nebulosity, 
the PSF fit converged for only 1 of the 28 sets of 
observations. Therefore, we report aperture magnitudes at H-band as well,
computed with an aperture radius of 4$''$ and a sky annulus 
extending radially from 14$''$-20$''$.  BN was not measured reliably at J-band.

Except for the time series aspect, the dataset, as produced by IPAC,
is identical in format to that produced for the 2MASS survey itself, 
containing position, photometry, photometry error, and photometric quality 
flags. To improve the photometric accuracy within the time series data
a grid of bright, isolated stars with low night-to-night variations 
was defined over the full $\sim0.84^\circ\times6.0^\circ$ survey region and
used as internal standards to adjust the nominal 
2MASS calibration zero points on a nightly basis. Typical zero point
corrections were $<$ 0.015 mag. The details of this procedure and all other
processing and analysis steps are described in Carpenter et al.

\section{Variability Characteristics of BN}

\begin{figure*}[tbp]
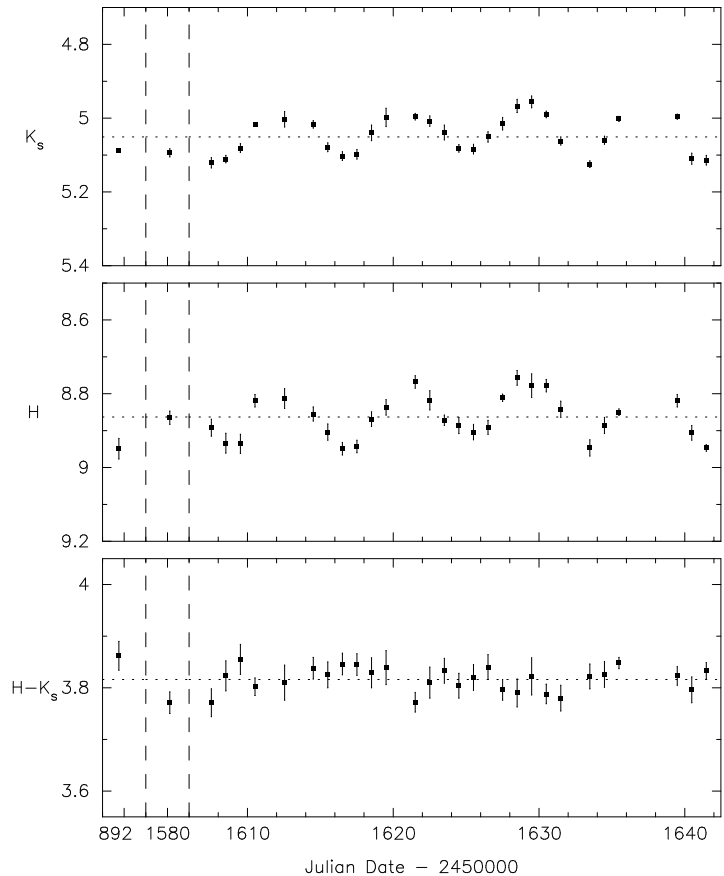

\insertplot{fig1.ps}{7.8}{8.4}{0.0}{0.8}{0.5}{0}
\caption{
Lightcurves in K$_s$-band, H-band, and H-K$_s$ color for BN 
as seen by the 2MASS southern telescope in 2000 March/April.
Dotted line is the mean magnitude over this time interval.  The first
two data points were taken during normal 2MASS survey operations 
and are disjoint from the rest of the data stream.
}
\label{fig:fig1}
\end{figure*}

For all of our $\sim$18,000 point sources, we looked
for photometric variability by comparing the observed
brightness changes in time to those expected according to the formal
photometric uncertainties.  To quantify the  likelihood that variability 
occurred within the timespan of the observations
we employed both a $\chi^2$ technique and
a method developed by Welch \& Stetson (1993) and Stetson (1996) 
which looks for correlated variability between multiple photometric bands. 

The BN measurements exhibited $\chi{_\nu}^2(K{_s})$ = 21, $\chi{_\nu}^2(H)$ = 13, 
and a Stetson-J variability index of 3.1 
(where we have considered values of $\chi_\nu^2 > 1.5$ and Stetson-J $>$ 0.55 
in Carpenter et al. to identify variables).  Lightcurves 
appear in Figure~\ref{fig:fig1} and the data in Table 1.
The mean magnitudes of BN during the 2000 March/April time period were 
H = 8.87 and K$_s$ = 5.04.  The error-weighted root-mean-squared 
of the measurements, which are proportional to the variability amplitudes,
were 0.06 mag at H and 0.04 mag at K$_s$ compared to typical photometric 
uncertainties of $<$0.02 mag.  The observed peak-to-peak amplitudes, 
neglecting errors in the photometry, 
were 0.26 at H, 0.17 at K$_s$, and 0.13 at H-K$_s$.

\begin{figure*}[tbp]
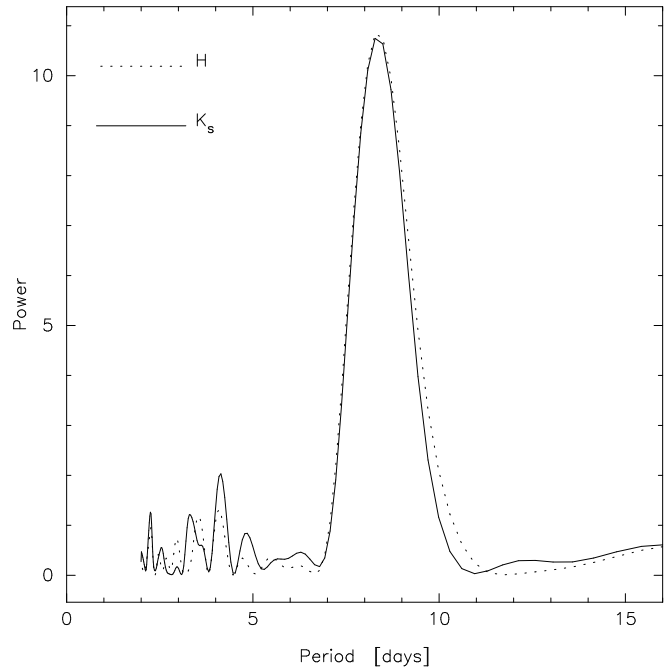

\insertplot{fig2.ps}{7.8}{8.4}{0.0}{0.8}{0.5}{0}
\caption{
Periodogram for H- and K$_s$-band data.  The most significant peak at 8.28 days
has a false alarm probability in the Lomb-Scargle formalism of $<$0.1\%. 
}
\label{fig:fig2}
\end{figure*}

\begin{figure*}[tbp]
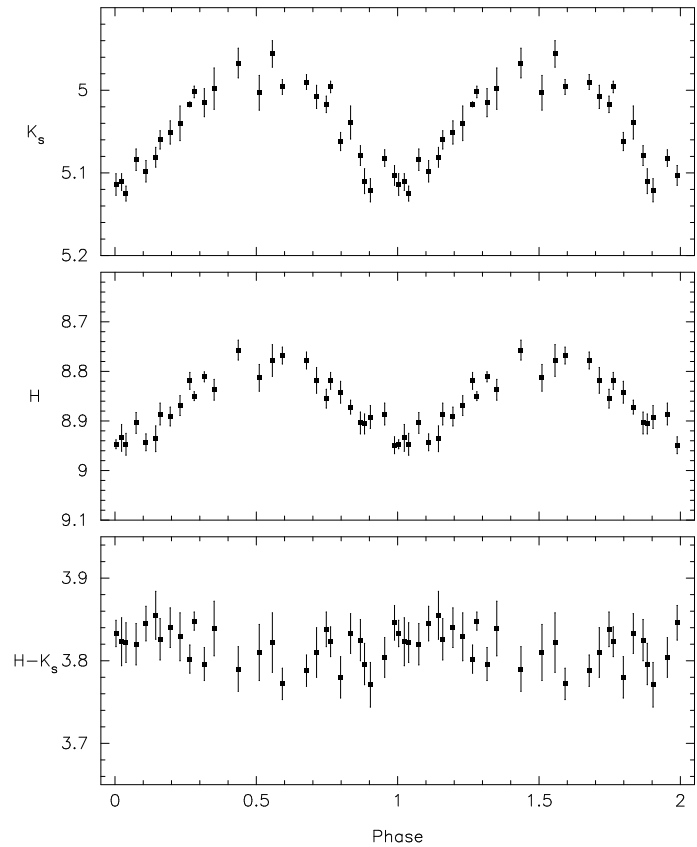

\insertplot{fig3.ps}{7.8}{8.4}{0.0}{1.0}{0.48}{0}
\caption{
K$_s$-band, H-band, and H-K$_s$ lightcurves phased with a period 
of 8.28 days.  Two full phases are shown. 
}
\label{fig:fig3}
\end{figure*}

For BN the lightcurve is clearly periodic, and application of the 
Lomb-Scargle algorithm from Press et al. (1992) yields a period of 
8.28 days at both H and K$_s$ (analyzed separately)
with false alarm probabilities (FAP) of $<$0.1\%.  
The error in this period according to the Kovacs (1981) formula for frequency
shifts in fourier analyses, is 0.05 days.
We show the periodograms in Figure~\ref{fig:fig2} and the phased lightcurves 
in Figure~\ref{fig:fig3}.  The H-K$_s$ color may also be periodic with the
same oscillation as the H and K$_s$ fluxes; however, 
this period is not significant
(FAP = 38\%).  Color variations are such that H-K$_s$ is redder
when the star is fainter, in the proportions expected from a standard
extinction law.  

Returning to Figure~\ref{fig:fig1}, in addition to the periodicity, 
there is a brightening of the magnitudes by $\sim$0.1 mag at H and 
$\sim$0.05 mag at K$_s$ over the first three cycles of the period which is
more apparent once the sinusoidal behavior is subtracted. Yet the faintest 
point at the end of the third cycle and the last two points in the data stream
at the end of the fourth cycle are all too faint to support that this
apparent rise is a long term behavior.
Further, the 2000 February and 1998 March flux levels indicate that
this recent brightening trend did not originate much before the beginning
of the 2000 March/April time series.

\section{Interpretation and Discussion}

Given the protostellar nature of BN, it is worth considering the physical
origin of the 1.6-2.2$\mu$m flux
(the shortest wavelengths at which BN has been detected).
Assuming the B3-B4 spectral type and 17 mag of visual extinction
from Gezari et al., the H-band magnitude matches within a few tenths 
that predicted for a reddened stellar photosphere. 
We take this as minor evidence that the H-band photometric variability 
may arise close to the photosphere.  At K$_s$-band, however, the (de-reddened) 
magnitude is almost 4 mag above the predicted photosphere.  The hot 
dust and/or gas producing the 2.2$\mu$m excess must subtend an area 
larger than that predicted by a standard blackbody disk 
or shell model since all close-in grains are destroyed by stellar heating.  
Accretional heating is one way to do this.  That the K$_s$-band magnitude is 
dominated by circumstellar flux suggests, alternately, that the variability may 
occur in the dust envelope.  Despite these differences between
H and K$_s$ in the ratio of non-photospheric to photospheric flux,
the observed periods are the same with the ratio of period amplitudes 
consistent with reddening.  

Periodic variable stars 
have been studied optically in the ONC region
by Herbst et al. (2000, and references therein), Stassun et al. (1999), and
Rebull (2001).  Almost all stars identified as periodic in these unbiased 
studies have been low mass, $<2 M_\odot$.  Only one 
periodic variable earlier than mid-G has been found, JW660 with a mid-B
spectral type and a mass $\sim 6 M_\odot$ (Hillenbrand 1997).  
The period of 6.15 days reported by Mandel \& Herbst (1991) has, 
perhaps notably, not been found again
in subsequent observing seasons (Herbst et al. 2000).
BN, by contrast, is a B3-B4 star with a mass $\sim 6-8 M_\odot$ (and a maximum
mass of 20 M$_\odot$ if the B0 spectral type inferred from HII region
characteristics is adopted).  It is the most massive periodic variable 
detected thus far in the Orion region. The next brightest stars 
having significant periods in Carpenter et al. are $>$2.2 mag 
fainter at K$_s$ with spectral types K0 and later. 

BN's periodic variability could be due to a number of well-recognized 
phenomena, although none of the following explanations 
seems totally satisfactory.

Periodic variability in young stars is usually interpreted in terms of
long-lived nonuniformities in photospheric structure, i.e. spots that are either
cooler or hotter the stellar effective temperature.  These spots 
rotate with the star and modulate the lightcurve as they pass through
the line of sight of the observer.  
Interpreted as stellar rotation, the 8.28 day period implies an equatorial 
velocity of $\sim$30 kms$^{-1}$ which is on the slow tail
for rotation of intermediate- and high-mass young stars 
in the ONC (Wolff, Strom, \& Hillenbrand 2001).  It is not generally 
accepted that massive stars like BN have the surface magnetic structures 
required for production of cool spots.  However, given that BN is
a protostar, accretion may produce surface shocks perhaps also requiring
magnetic fields in the form of ordered magnetic dipoles which lead to
hot spots as material from the circumstellar environment falls in along them.

Considered independently,
the variability amplitudes at H and K$_s$ can be well-modelled by spots 
with $\Delta$T$\approx$15,000 K from the photosphere and $\sim$15-20\% 
coverage, or $\Delta$T$\approx$5,000 K and $\sim$50\% coverage, as examples.
However, neither cool nor hot spots are capable of producing the small 
H-K$_s$ color amplitude since the effect of adding a spot
is essentially colorless in the near-infrared given the early B photosphere  
($<$0.02 mag for $\Delta$T$<$20,000 K and coverage $<$50\%).
If spot-modulated rotation is the cause of the observed photometric
periodicity, the phase and the amplitude both should change
on timescales of months to years as the spot structure varies.  
Although the 2000 February data point
phases well with the period derived for the 2000 March/April time series, 
the 1998 March data point does not; however, this could be due simply 
to accumulation of period error over the longer time baseline.


Pulsating behavior leads to short periods (0.1-0.3 day) in radial modes 
and to only slightly longer periods ($<$1 day) in non-radial modes 
(e.g. $\beta$Cep stars in the
early B range and 53 Per type stars and others at late B types).  
Both the amplitude
and the near-sinusoidal shape of BN's lightcurve are consistent with certain
types of pulsational behavior, but the period is too long.  
Longer period variability ($\sim$2-30 days) in massive stars is often explained 
in terms of winds, with some mechanisms requiring binary systems.  
The radio spectral index of BN is $\alpha=0.8\pm0.2$ from 2-6 cm
(Felli et al. 1993), consistent with thermal emission from an 
ionized stellar wind.  

The observed period and mass estimate for BN imply an orbital radius 
of $\sim$0.15 AU for any hypothetical low-mass companion.
If an outflow/wind from a companion was colliding with 
the outflow/wind from BN, pulsating behavior within the interaction
region might be identified via
x-rays (e.g. Ishibashi et al. 2000 for Eta Car). BN was seen in heavily 
absorbed x-rays by Chandra (Garmire et al. 2000).  However, the Chandra 
position was 1.1" northwest ($\sim$500 AU) of the near-infrared position, 
whereas all other x-ray/optical-infrared matches were within 0.5".  
Garmire et al.  attribute the offset, if real, to physics associated
with outflow phenomena interacting with stationary, dense cloud material;  
no binary system is required. 

Another scenario to consider is an eclipsing binary.  
BN's lightcurve does not seem consistent with 
a fully eclipsing system given its low amplitude and nearly sinusoidal nature.  
A partial eclipse situation with a near-equal mass/size companion,
a small orbital separation ($\sim$10 R$_*$), and a reasonable inclination 
($\sim40^\circ$) would match the gross shape, period, and amplitude 
of the phased lightcurves.
In the near-equal mass situation, the true orbital period would be double 
that derived naively from observations since a single revolution 
produces two minima and two maxima.  However, eclipses should exhibit 
the same amplitude in all bands (modulo limb darkening), which is not 
what is observed for BN unless the companion has the same radius but the colors
of a much cooler star.  Interestingly, Scoville et al. (1983) and others have
suggested the possibility of wide binarity for BN to explain its large 
radial velocity relative to other ONC stars and to the ambient cloud.

A further possibility is periodic occultation by asymmetry 
in a circumstellar disk at the 0.15 AU orbital radius implied by the period.  
High column density, partially grey, orbiting material could produce 
the shape of the lightcurve as well as the color and magnitude amplitudes. 
It is interesting in this context to note that
Biscaya et al. (1997) claimed time variability in BN's 2$\mu$m CO bandhead 
emission lines, and that no emission is present in the spectrum of 
Penston et al. (1971).  The CO lines are formed in the dense, hot circumstellar
environment and their time variability may have the same 
physical origin as the continuum variability found by us.

Finally, we note that there may be historical precedent for photometric
variability in BN, as detailed in Table 2.  One must be cautious
in interpreting this ensemble of early near-infrared measurements 
as indicative of large scale flux variations, however.  The 
observational complexities of working in
the Orion region combined with a variety of aperture sizes
may explain entirely the apparent differences in photometry. 

\section{Summary}

We have found photometric modulation in H- and K$_s$-band lightcurves for 
BN consistent with periodic behavior.  During 2000 March/April 
the period was 8.28 days with a Lomb-Scargle 
false alarm probability of $<$0.1\%.  The amplitude of the nearly-sinusoidal
lightcurve was $\sim$0.2 mag peak-to-peak ($\sim$0.05 mag root-mean-squared).  
The origin of the periodicity is not immediately
obvious.  Modulation of the lightcurve due to rotation of inhomogeneities in
either the photosphere or the inner circumstellar dust distribution
is the least complicated model.
Further multiwavelength photometric as well as emission-line profile monitoring
can probe period persistence, phase stability, and physical origins for
the periodic behavior.

\acknowledgements

This publication makes use of data products from 2MASS, 
which is funded by NASA and NSF and
is a joint project of the University of Massachusetts and 
the Infrared Processing and Analysis Center at Caltech. 
Science data and information services were provided by IRSA at IPAC.
The authors would like to thank the 2MASS Observatory Staff
and Data Management Team for acquiring and pipeline processing
the special survey observations used in this investigation.
We thank Nick Scoville, Bill Herbst, Michael Meyer, and Roc Cutri 
for comments on drafts of this manuscript.  We also thank Gerry Neugebauer
for communicating previously unpublished data on BN. 


\begin{table*}
\caption{Time Series Photometry for the BN Object}
\begin{tabular}{rrrrr}
\noalign{\smallskip}
\hline\hline
Julian Date   &    H   &   K$_s$  & err(H)& err(K$_s$) \\
               & [mag]  & [mag]    & [mag] & [mag] \\
\hline
2450891.509 &  8.949 &5.087 &0.028& 0.004 \\
2451580.546 &  8.865 &5.094 &0.018& 0.011 \\
2451607.526 &  8.892 &5.121 &0.023& 0.014 \\
2451608.520 &  8.934 &5.111 &0.027& 0.010 \\
2451609.520 &  8.936 &5.081 &0.026& 0.012 \\
2451610.519 &  8.819 &5.017 &0.017& 0.003 \\
2451612.553 &  8.813 &5.003 &0.027& 0.021 \\
2451614.516 &  8.855 &5.017 &0.019& 0.010 \\
2451615.515 &  8.904 &5.079 &0.022& 0.012 \\
2451616.514 &  8.949 &5.103 &0.017& 0.012 \\
2451617.514 &  8.943 &5.098 &0.017& 0.013 \\
2451618.513 &  8.869 &5.040 &0.020& 0.021 \\
2451619.511 &  8.837 &4.998 &0.021& 0.025 \\
2451621.510 &  8.768 &4.996 &0.017& 0.009 \\
2451622.512 &  8.818 &5.008 &0.026& 0.014  \\
2451623.509 &  8.872 &5.039 &0.014& 0.020  \\
2451624.507 &  8.886 &5.082 &0.022& 0.010  \\
2451625.507 &  8.904 &5.084 &0.021& 0.013  \\
2451626.506 &  8.891 &5.051 &0.019& 0.014  \\
2451627.505 &  8.811 &5.015 &0.010& 0.017  \\
2451628.499 &  8.757 &4.967 &0.020& 0.018  \\
2451629.498 &  8.778 &4.956 &0.032& 0.016  \\
2451630.498 &  8.778 &4.990 &0.017& 0.009  \\
2451631.497 &  8.842 &5.062 &0.022& 0.011  \\
2451633.491 &  8.947 &5.125 &0.022& 0.009  \\
2451634.490 &  8.886 &5.060 &0.022& 0.011  \\
2451635.489 &  8.850 &5.002 &0.009& 0.007  \\
2451639.485 &  8.819 &4.996 &0.017& 0.007  \\
2451640.480 &  8.906 &5.110 &0.020& 0.015  \\
2451641.479 &  8.947 &5.114 &0.009& 0.013 \\ 
\hline
\hline
\end{tabular}
\end{table*}

\begin{table*}
\caption{Historical Photometry for the BN Object}
\begin{tabular}{cccllcc}
\noalign{\smallskip}
\hline\hline
Reference &Observation Date    &Aperture Size  &  H   &   K  & err(H)& err(K)\\
          & [UT]               & [arcsec]      &[mag]  & [mag]&[mag]  & [mag]\\
\hline
Becklin \& Neugebauer (1967) &1965, January&13   &9.8 &5.2  & ? & ? \\
Neugebauer (2000)            &1968, September 15&?&9.19&4.88&0.10&0.07\\
Low et al. (1970)            &   ?    &?    &9.60&4.87 & ? & ? \\
Neugebauer (2000)            &1969, December 8&? &8.57&4.72 &0.15&0.15\\
Penston (1973)               &1971, March 9&15   &8.48&4.5  &0.13& ?\\
Neugebauer (2000)            &1974, September 21&7&9.45&4.76&0.10&0.10\\
Lonsdale et al. (1982)       &1980, February&3.5  &9.2 &5.1  &$<$0.3&$<$0.3\\
Neugebauer (2000)            &1981, March 14&6&9.39&4.93&0.06&0.04\\
Hyland et al. (1984)         &1982 or 1983  &4    &--  &5.5  &-- &$<$0.01\\
Minchin et al. (1991)        &1988, January&6    &9.8 &5.4  &0.1&0.1\\
This Paper                    &1998, March 19&8    &8.95&5.09 &0.03&$<$0.01 \\
Hillenbrand \& Carpenter (2000)&1999, February 9&1.8&9.36&--   &$<$0.01&-- \\
This Paper, mean over time series&2000, March/April&8    &8.87&5.04 &$<$0.03 &$<$0.02 \\

\hline
\hline
\end{tabular}
\end{table*}

\clearpage

%
%

\end{document}